\documentclass[epsfig,12pt]{article}
\usepackage{epsfig}
\usepackage{graphicx}
\usepackage{geometry}
\usepackage{array}
\usepackage{color}
\usepackage{bm}
\usepackage{hyperref}
\hypersetup{
	colorlinks=true, 
	linktoc=page,
	linkcolor=blue}
\usepackage[nottoc,notlof,notlot]{tocbibind} 
\usepackage[titles]{tocloft}

\newcommand{\beq}{\begin{equation}}   
\newcommand{\eeq}{\end{equation}}
\newcommand{\beqn}{\begin{eqnarray}}   
\newcommand{\eeqn}{\end{eqnarray}}

\usepackage{amsmath,amssymb}

\newcommand*\DAlambert{\mathop{}\!\mathbin\Box}

\newcommand{\gsim}{\lower.7ex\hbox{$
\;\stackrel{\textstyle>}{\sim}\;$}}
\newcommand{\lsim}{\lower.7ex\hbox{$
\;\stackrel{\textstyle<}{\sim}\;$}}
\begin{document}

\begin{flushright}

December 31, 2019

{\small FTPI-MINN-19/29,  UMN-TH-3908/19}

\end{flushright}

\begin{center}
{\large  MUSINGS ON THE CURRENT STATUS OF HEP}

\vspace{2mm}

M. SHIFMAN

\vspace{2mm}

{\em \small William I. Fine Theoretical Physics Institute, 
University of Minnesota,
 Minneapolis, MN 55455, USA}

\end{center}

In 2012 I published an article \cite{impres} presenting my personal views on the state of affairs in ``our science,''  HEP theory.\footnote{HEP is the abbreviation for High Energy Physics.} That ``old'' article opens with a rather optimistic epigraph ``Paraphrasing Feynman: Nature is more  imaginative than any of us and all of us  taken together. Thank god, it keeps sending   messages rich on  surprises." Has Feynman's prophecy come true?

2012 was also the year of the Higgs boson discovery closing the age of the Standard Model (SM) confirmation. 

 Now, seven years later,  I will risk to offer my musings on the same subject. The seven years that have elapsed since \cite{impres} brought new perspectives:
the tendencies which  were rather foggy at that time became pronounced. My humble musings do not pretend to be more than they are: just a personal opinion of a theoretical physicist... For obvious reasons I will focus mostly on HEP, making a few marginal remarks on related areas. 

I would say that the most important message we have received is the absence of dramatic or surprising new results. In HEP no significant experimental findings were reported,\footnote{This statement does {\em not} refer  
to {\em astrophysics} and {\em cosmology}.} old ideas concerning Beyond the Standard Model (BSM) physics hit dead-ends one after another and were not replaced by novel ideas. Hopes for key  discoveries at the LHC (such as superpartners) which I mentioned in 2012 are fading away. Some may even say that these hopes are already dead. 
 Low energy-supersymmetry is ruled out, and gone with it is the concept of naturalness, a basic principle\,\footnote{By the way, this principle has never been substantiated by arguments other than aesthetical.} which theorists cherished and followed  for decades. Nothing has replaced it so far.\footnote{An alternative -- Multiverse in conjunction with  anthropic principle -- go beyond the conventional paradigm of physics as a natural science. I will  discuss it in brief on page  \pageref{p14}.} With the disappearance of this principle the issue of mass hierarchies becomes almost (if not completely) meaningless.\footnote{The same refers to fine-tuning of $\theta$, see below.} 
The Standard Model is still unchallenged: today no observed natural phenomena require its expansion.  Dark matter composition is still a huge question mark.
With a few exceptions (in quantum field theory at strong coupling), expectations of breakthrough developments in HEP theory and related areas did not materialize.

Of course we could, and should, rejoice with the colleagues working on the mysteries of our Universe -- their work was rewarded by the recent discovery of gravitational waves  predicted by Einstein
103 years ago.\footnote{ In fact, the situation with the gravitational wave prediction was more complicated. In 1936 Einstein wrote to
 Max Born:
``Together with a young collaborator [Natan Rosen], I arrived at the interesting result that gravitational
waves do not exist, though they had been assumed a certainty to the first approximation."
Their paper was submitted  to the Physical Review
under the title ``Do Gravitational Waves Exist?'' Einstein and Rosen (ER)
rediscovered the so-called Beck vacua \cite{beck}, a family of gravitational wave solutions with cylindrical symmetry. While analyzing them
ER misinterpreted a coordinate singularity in their solutions as instability. ER's manuscript was retracted from Physical Review shortly after but the debate dragged for decades, till the mid-1950s when the so-called ``Sticky bead argument" was anonymously put forward by Richard Feynman at a conference at Chapel Hill. Here is how it was formulated by Feynman in a private letter \cite{RPF}: 	``Feynman's gravitational wave detector: It is simply two beads sliding freely (but with a small amount of friction) on a rigid rod. As the wave passes over the rod, atomic forces hold the length of the rod fixed, but the proper distance between the two beads oscillates. Thus, the beads rub against the rod, dissipating heat."} Moreover, the discovery of a nonvanishing cosmological  constant (CC) a decade ago or so,
\beq
\rho_{\rm vac.\, density} \sim 10^{-47} {\rm GeV}^4 \sim \big( 2\times 10^{-3} {\rm eV}\big)^4
\eeq
led to a dramatic change of  a paradigm. Previously theorists were aimed at explaining the vanishing of CC by virtue of a symmetry. It is much harder to understand why
 CC$\neq 0$ but is so incredibly small.
Fundamental discoveries in astrophysics and cosmology continue, which make physicists working in this area happy. 

But this is not the area in which I work. HEP, ``my" branch of theoretical physics since the beginning of my career, seems to be shrinking. A change of priorities in HEP in the near future is likely as business as usual is not sustainable. The current time is formative. 

Such turn of events is by no means unique. Classical physics which flourished for centuries gave place to quantum physics in the very beginning of the 20th century.
The difference is that {\em  then} the experimental data forced theoretical physicists to switch to a new quantum paradigm. What should happen for today's HEP theory to reincarnate itself?   It is not clear to me. It seems
that I see a renewed interest in this endeavor among bright young people. Hopefully, it is not wishful thinking. 

Meanwhile, more traditional  HEP physicists do not hibernate. The routine work goes on unabated, people work hard to polish the ideas that had been put forward previously.
Theorists revisit corners which were ignored on the previous journeys.

In 2012 I wrote that theorists' MO\,\footnote{Modus Operandi.} could be called (somewhat conditionally) ``the giant resonance mode. In this mode each novel idea, once it appears, spreads  in an explosive manner in the theoretical community, sucking into itself a majority of active theorists, especially young theorists. Naturally, alternative lines of thought by and large dry out. Then, before this given idea brings fruits in understanding phenomena occurring in nature (both, due to the lack of experimental data and due to the fact that on the theory side 
crucial difficult problems are left behind, unsolved), a new novel idea arrives, the old one is abandoned, and a new majority jumps onto the new train.''  The outstanding mathematician Alain Connes once wrote:

\begin{quotation}

In general mathematicians tend to behave like
 ``fermions" i.e. avoid working in areas which are too trendy whereas physicists behave a lot more like ``bosons" which coalesce in large packs and are often ``overselling" their doings, an attitude which mathematicians despise \cite{connes}. 
 
 \end{quotation}
 \noindent
It is quite possible that 
the lack of new ideas  we are currently witnessing will make the HEP community switch to the ``fermion-like" MO. If so, this  will be the first clear-cut response to today's  challenges.

\vspace{2mm}

\centerline{***}

I will start my musings by sketching  a huge quantum tree which grew out of the discovery of quantum mechanics in the 1920s. The reason for this digression is two-fold.
First, I would like to explain why both HEP  and condensed matter theory experienced radical changes in the 1970s and say a few words 
about their second lives today.  Second, I will argue that developing parts of HEP today are in fact the same as developing parts of quantum field theory.

\begin{figure}[h]
\epsfxsize=10cm
\centerline{\epsfbox{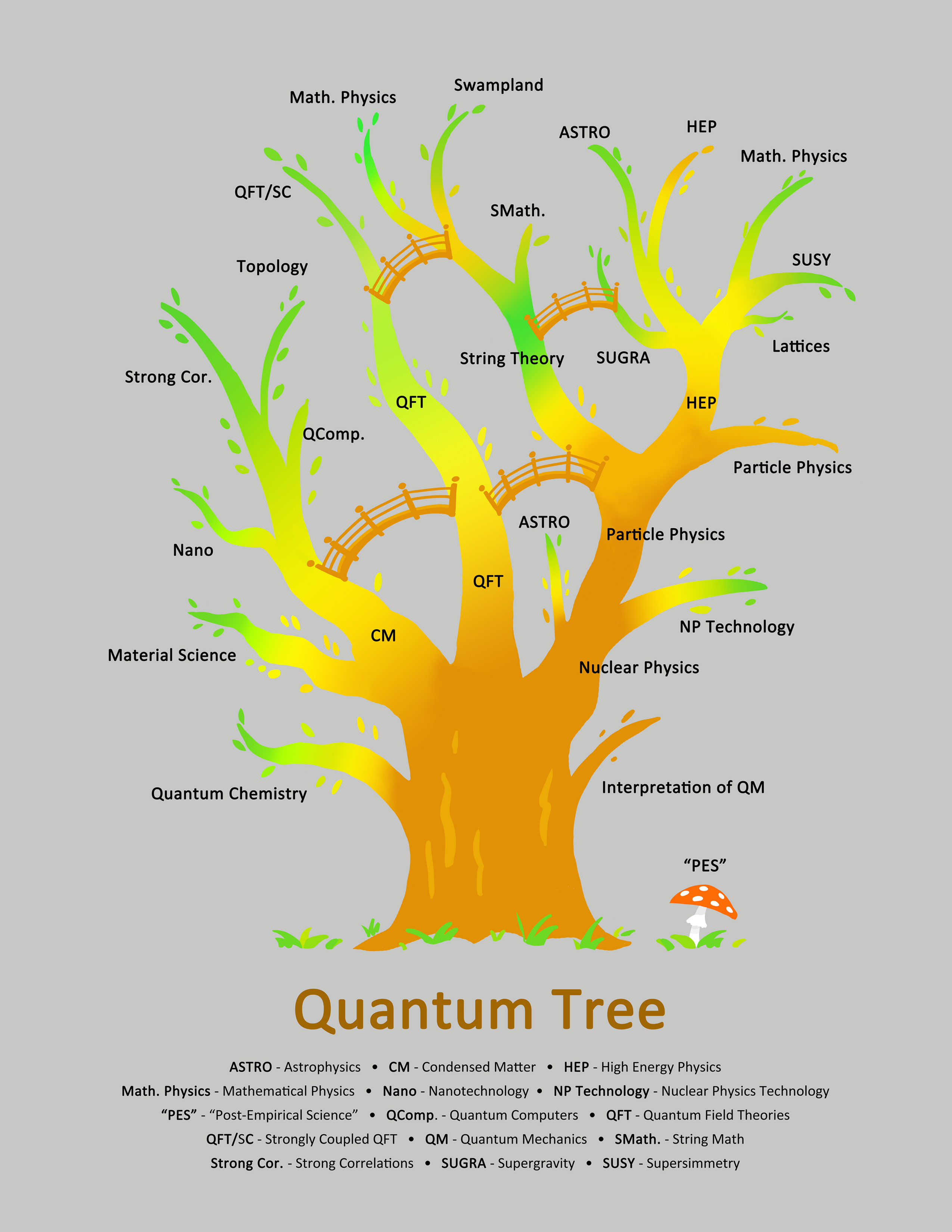}}
\vspace{-1.5mm}
\caption{\small
Quantum tree. The growing branches are green.  }
\label{figu3}
\end{figure}

In Fig. \ref{figu3}
I  present  a simplified picture of the Quantum Tree. The quantum story begins in the early 1920s when  quantum mechanics was discovered. Three 
earliest branches which grew on the tree were condensed matter (CM), quantum field theory (QFT) and nuclear physics (NP). The first  branch out of three listed above produced quantum chemistry, material science and modern CM which after 1970 acquired  fresh branches, e.g.  nanophysics. The third branch gave rise to particle physics (PP) which later transformed itself into HEP and astroparticle physics (AP).

At the third level, at the top of the quantum tree, we see modern disciplines: quantum information/computing, physics of strongly correlated matter, string theory, string mathematics, supersymmetry and supergravity, strongly coupled QFT, and a few others. 

 With time some of the old branches died or nearly died out as scientific disciplines. 
For instance, what was nuclear physics in the 1930s -- 1950s in part reincarnated itself as nuclear technology. Its other part fused with HEP and AP.
Particle physics and HEP gave birth to string theory. At birth the baby was christened ``Veneziano amplitude" \cite{va}. It grew into a powerful branch which made many believe  
that the ``theory of everything" is around the corner. Well... it never happened and -- I will risk to say -- never will. At the same time people harvested precious fruits from the string/brane branch, for instance,  gauge-gravity duality, holography, and many qualitative insights in Yang-Mills dynamics at strong coupling in supersymmetric QFT.

From Fig.~\ref{figu3}  you can see that the PP branch is currently withering. HEP and ST  were rapidly expanding since the 1970s until approximately 2000 or so, then this growth flattened off and the tendency reversed itself. HEP redefined itself as quantum field theory at strong coupling (including supersymmetric methods and tools), with some islands of phenomenology here and there. The string theory sprout which is still ``work in progress"  is called rather unconventionally -- ``Swampland" \cite{swamp, swamp1},  see the top of the tree in Fig. \ref{figu3}. 

Courageous people reaching the very summit of ST  probably reason as follows: ``OK, string theory is too complicated and too distant from our today's world knowledge so that currently we cannot  fill in all gaps making it a complete theory of everything in our world. Moreover, string Landscape is so vast... However, maybe, we can analyze its most general features and infer which QFT classes might emerge at energies much below the Planck scale and become valid candidates for our world. Those which cannot,  lie in the swampland and are not worth consideration." Similar ideas come from {\em gedanken} experiments with black holes.


\begin{figure}[h]
\epsfxsize=10cm
\centerline{\epsfbox{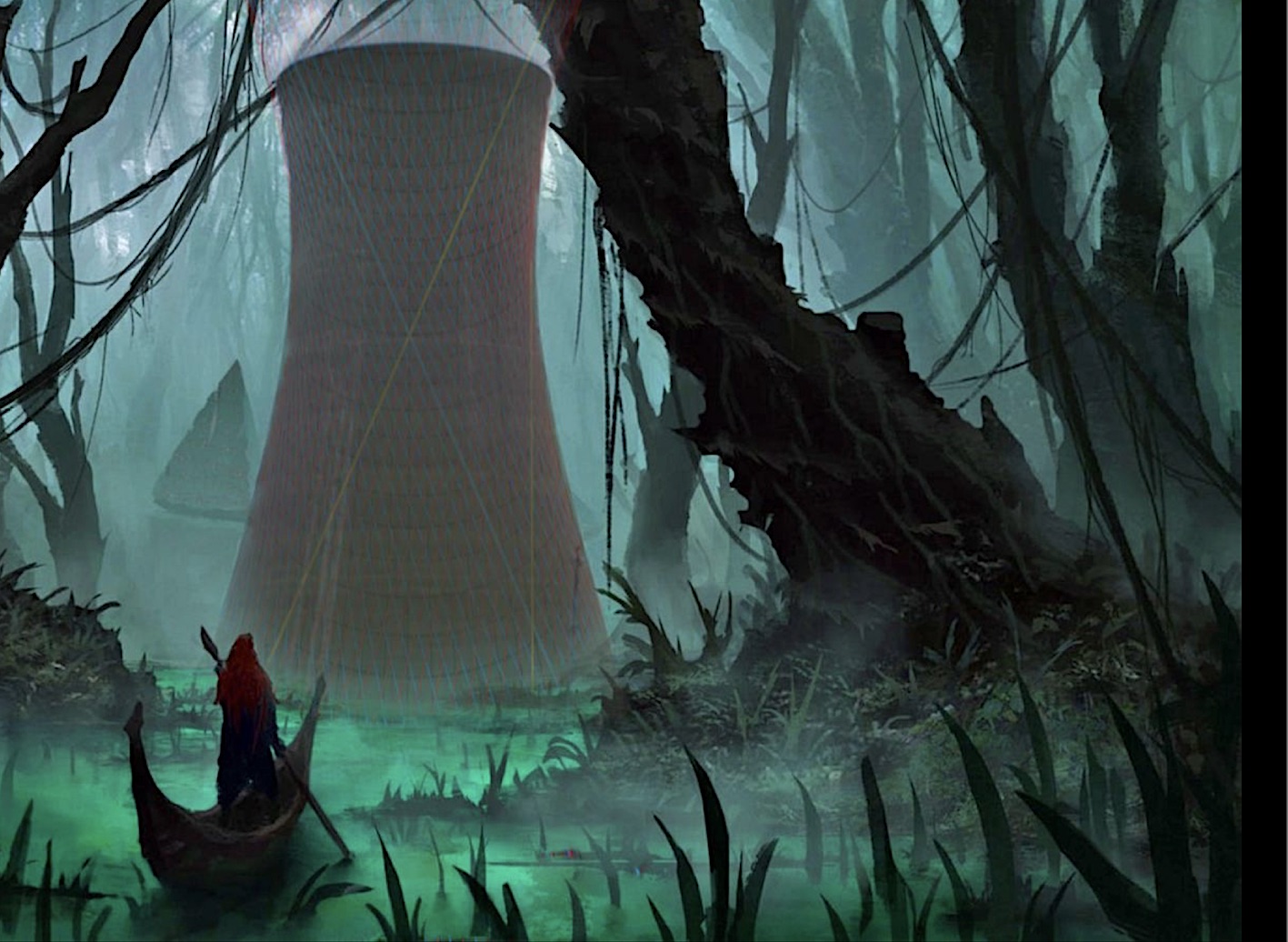}}
\caption{\small
Swampland (courtesy of IFT, Madrid).}
\label{figu1o}
\end{figure}

One relatively simple example of this type is the so called ``Weak Gravity" \cite{wg}. According to this argument, no particles can exist whose electric charges are so small that
their electromagnetic interaction is weaker than gravitational interaction. In other words, gravity is the weakest long-range force.\footnote{In our world this statement was derived long ago from other arguments \cite{OVZ}.} 

The theorists involved went far beyond exploring our world.  One can say that today's theorists mainly investigate imaginary worlds, the worlds  which might have existed as an alternative to our world being somewhat similar to ours.\footnote{This characterization was suggested by Andrei  Losev.} The degree of similarity may vary, from very similar to our  world (e.g. changing the number of colors and space-time dimensions is quite fruitful, so we are happy) to mildly fantastic (e.g. adding supersymmetry), to those which  -- I am afraid -- could  be characterized as 	``El sue\~no de la raz\'on produce monstruos.'' Whether it is good or bad --  time will show.

\begin{figure}[h]
\epsfxsize=10cm
\vspace{-2mm}
\centerline{\epsfbox{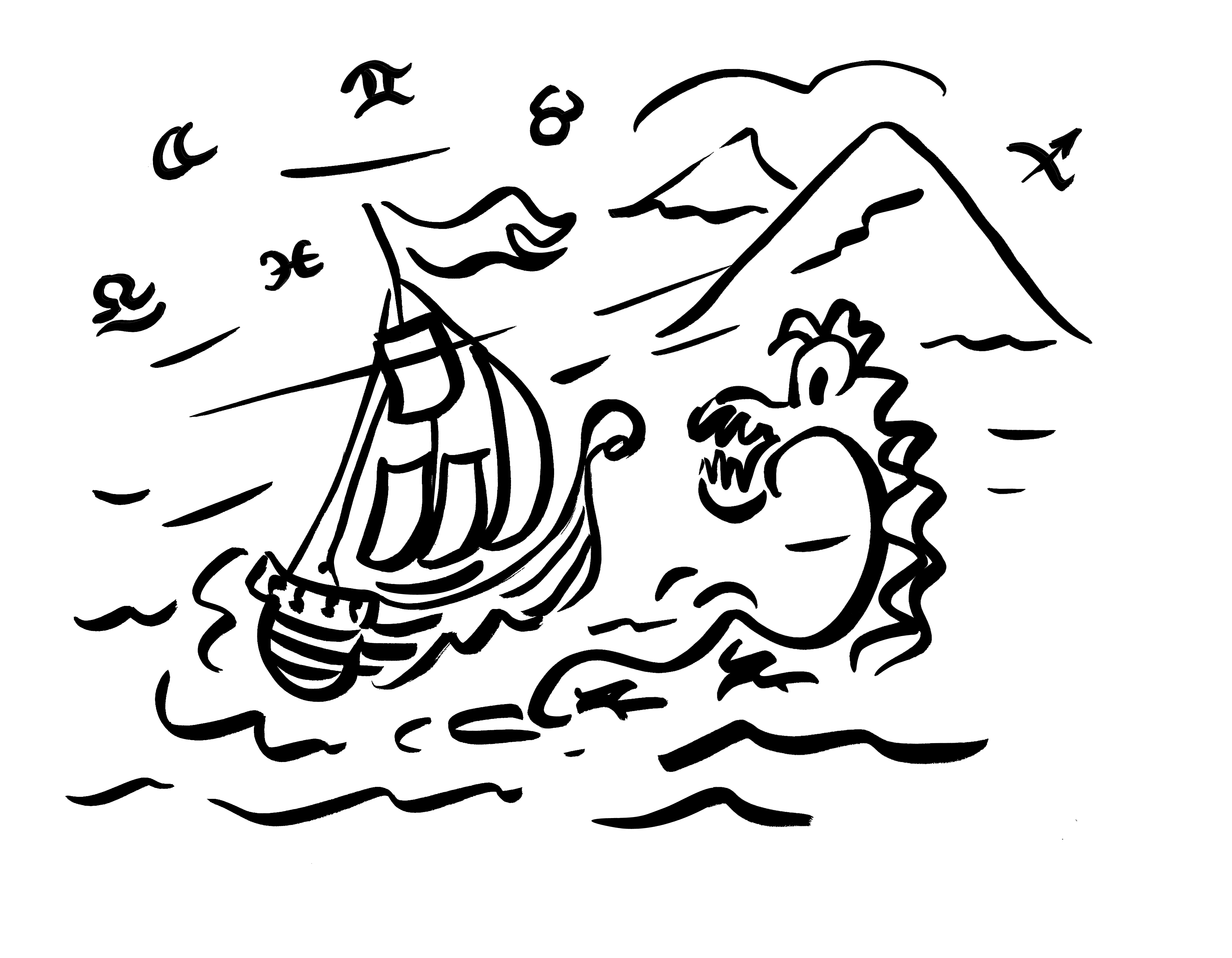}}
\vspace{-6mm}
\caption{\small
A mildly fantastic scenario? }
\label{figu1ok}
\end{figure}

A few words are in order concerning a sprout which branched off from ST branch -- string mathematics, or SMath as it is labeled in Fig. \ref{figu3}. I think that in the future it will fuse with \label{p6} the
Mathematics Tree (not shown in this figure). Mathematical questions emerging from ST are subject to the same logic as mathematics at large. Although historically mathematics developed out of practical needs, at present its philosophy and methodology are drastically different than that of theoretical physics. In math it is customary to start from a set of postulates (axioms) and rigorously derive as many applications (theorems) as possible. {\em En route} mathematicians establish whether the initial set was complete and selfconsistent. Math is about constructions (quite often, beautiful). Whether or not they may be used for description of nature is secondary.

Two green branches at the top in Fig.~\ref{figu3} labeled ``Math. Physics" show that its development is steady.

Small bridges in Fig. \ref{figu3} indicate interconnections between the branches of the tree. I draw a few of them and anticipate this cartoon to be severely criticized by many readers, and justly so. Indeed, 
it is hard to imagine drawing some such figure which would be complete and historically faithful since the number of connections between ideas is huge.   
Moreover,  QFT is not a branch like others, a subject of its own. It is a framework that underlies a lot of cosmology, a lot  of CM, and all of particle physics. 
 I sinfully draw it as a branch connected to the rest by only three small bridges. In my defense I can only say that this is the best I could do in planar geometry. There are many more interconnections at every stage and in every direction. The reader will have to use his/her imagination to visualize them.
 
 \vspace{2mm}

\centerline{***}

In the 1970s (this time roughly corresponds to the middle level of the tree in Fig.~\ref{figu3}), CM theory which  had been  previously based essentially on quasiparticle description and quantum mechanics shifted toward QFT. New key words appeared in CM vocabulary: universality class, topological state of matter, path integrals, etc. They were borrowed from QFT.

In its turn QFT obtained a second life after the discoveries of 
asymptotic freedom,  supersymmetry and supergravity. These fundamental shifts led to a powerful growth of the tree in all directions, as is seen from this figure. Quantum chromodynamics (QCD) was firmly established. Approximately at the same time, after the discovery of the $c$ quark and $\tau$ lepton, the Glashow-Weinberg-Salam model of electroweak interactions evolved in the Standard Model. This was the triumph of HEP, a success achieved because theory and experiment went hand in hand with each other being powered by each other.
A remarkably thorough understanding of empiric data accumulated by this time was achieved. Theorists worked with joy and enthusiasm, all disconnected pieces suddenly came together and -- within a decade -- conceptual questions on strong and electroweak interactions were understood and answered. I was lucky that my professional career started in 1973. Till now I vividly remember the stormy days of the ``November revolution" in 1974. The few months following the discovery of $J/\psi$ were the star days of QCD and probably the highest emotional peak in my career.

Now, let us discuss the current status of HEP. What was going on in HEP in the last seven years or so? To summarize the answer, on page \pageref{figu2ok} I present a cartoon made in 2010 on occasion of 
a conference dedicated to M.~Gell-Mann's 80th anniversary. Please, do not pay attention to geographical background: it was chosen arbitrarily and is unrelated to physics contents. 

What is important is that the areas around the dashed pink arrow are becoming depopulated. The reason is that the minimal supersymmetric standard model (MSSM) no longer seems relevant, as well 
as the very idea of low-energy supersymmetry
which was put forward  to solve the hierarchy problem. Basically, experimental data from CERN (or, better to say, their absence) ruled MSSM out. The concept of naturalness seemingly lost its appeal. 
By the way, if so, there is no need in the celebrated axion (see point 9 in the Map) to guarantee CP conservation in strong interactions. Indeed, without naturalness it could well happen that the $\theta$ angle was set very  close to zero by the same mechanism which fine-tuned the Higgs mass. There are no fresh ideas beyond SM either, with the exception of a few contrived, baroque and -- most probably -- unviable constructions suggested {\em ad hoc}. 
Can ongoing research in neutrino physics give us a hint?

On the other hand, explorations at the periphery of the Map continue. This is especially true with regards to the dark matter  (DM) mystery. A quick
 glance at the current HEP theory literature  is sufficient to verify this statement. The existence of DM is confirmed beyond any doubt. In fact, DM constitutes about $\frac 14$ of our Universe.
 But its composition remains unknown. Ten or 15 years ago the general belief was that DM is built of LSP -- the lightest supersymmetric particles. One of  the less popular alternatives which was under consideration 
 is the axion cloud around each galaxy. At present, one can find in the literature a spectrum of other hypotheses. Frankly, none of the newcomers seems aesthetically appealing. Nor are they motivated.
 What is good is that there is still hope that DM could be experimentally detected and studied. One should not forget, however, of a  scenario in which DM components interact with us only gravitationally. This scenario is not ruled out, and if it is realized in nature,  the DM structure will remain unknown in the near future.
 
 \vspace{1mm}
 \begin{figure}[h]
\epsfxsize=12.5cm
\centerline{\epsfbox{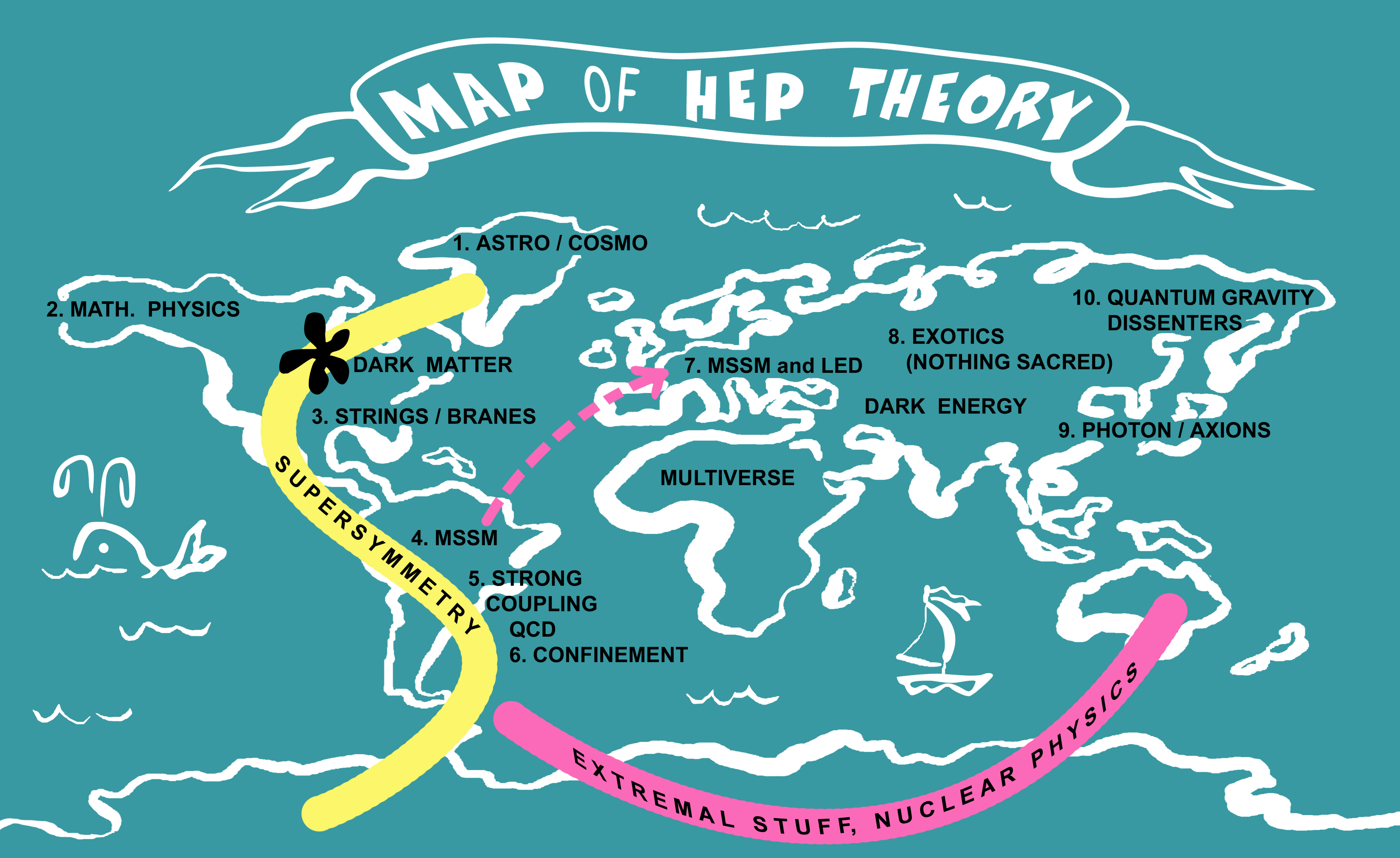}}
\caption{\small
 Map of HEP.}
\label{figu2ok}
\end{figure}

\vspace{2mm}

\centerline{***}

The theme of supersymmetry (SUSY) which is so conspicuous in the Map (Fig.~$\!$\ref{figu1ok})
redefined itself. To my mind, it is no longer dominated by SUSY phenomenology. The second face of SUSY --  supersymmetry as a tool for exploring gauge dynamics at strong coupling -- which started emerging 
in the early 1980s \cite{NSVZ} is taking  precedence over phenomenology. It is unclear for how long, though. 

A {\em decisive breakthrough} in this direction occurred in 1994, with the discovery of the Seiberg duality and the Seiberg-Witten solution of 
${\cal N}=2$ supersymmetric Yang-Mills theory slightly deformed by a mass term of the adjoint chiral superfield \cite{sei}. 
This remarkable achievemnt to a large extent feeds continuous advances of two branches at the top of the Quantum Tree in Fig.~\ref{figu3}: one labeled ``SUSY" on the right-hand side and the other labeled ``QFT/SC" on the left-hand side (SC stands for strong coupling). In fact, they are interconnected, but I could not visualize this connection in the planar cartoon.  QFT/SC is also intimately connected with the modern condensed matter theory. 
To mention just a few findings on the SUSY branch let me mention the exact 2D-4D correspondence (related to the Seiberg-Witten results) and non-Abelian vortex strings revealing a wealth of sigma models on their world sheets with varying degree of supersymmetry, including chiral SUSY. Studies of the above two-dimensional sigma models are of interest on their own.

A few words are in order about non-supersymmetric section of QFT. I am delighted to see that after a long relatively separate existence,\footnote{I write here ``{\em relatively} separate." In fact they were always connected, suffice it to mention K. Wilson's and A. Polyakov's ideas. Physics is our common edifice, after all! } HEP and CM theories are moving towards each other, with growing cross-fertilization.

A recent success in {\em non}-supersymmetric theories at strong coupling is the discovery of {\em mixed anomalies} (say, 1-form vs. chiral anomaly, see Appendix on page \pageref{app}) for {\em global} symmetries \cite{K1,K2,K3}. Often they are referred to
as 't Hooft anomalies.\footnote{Gerard 't Hooft was the inventor of {\em anomaly matching}. This idea played a prominent role  in modern QFT.
All anomalies are  ``all-scale" phenomena, they have a UV face and an IR  face which must match \cite{shia}. 
 't Hooft was the first to discus this aspect  in \cite{tho}.
The 't Hooft matching became extremely popular after  Seiberg's pioneering exploitation 
of supersymmetric QCD anomaly matchings (see the first reference in \cite{sei}).  Shortly after, the 't Hooft matching 
became a tool ``for everyone."
Everyone calls the mixed anomaly 't~Hooft
regardless of its 0-form or higher-form symmetry.}

The basic idea is that in some theories there are two global symmetries, which clash with each other at  the quantum level. Only one of them can be maintained. This provides unique information on the infrared (IR) behavior of the theory, in particular, its vacuum structure. In the past anomalies in global symmetries have not been considered in this perspective.\footnote{At least, not considered systematically, see, however, \cite{ss} for precursors.} Although they could have been uncovered much earlier, the phenomenon was overlooked. The simplest  pedagogical example is as follows.

Assume we consider two-dimenional Schwinger model with one massless Dirac fermion of charge 2 \cite{da1}. More exactly, in addition to the dynamical charge-2 fermion, there is a heavy probe charge-1 fermion
whose mass can be viewed as tending to infinity. Next, assume that  in this model we compactify the spatial dimension on a circle of circumference $L$, i.e. impose either periodic or antiperiodic boundary conditions on the fermion fields. Then one can show that this model has two discrete $Z_2$ symmetries -- one 0-form and another 1-form.
These  two global $Z_2$ symmetries have generators which do {\em not} commute with each other \cite{da1}. Thus, only one of these symmetries can be implemented, the other one must be spontaneously broken. Hence, the ground state is {\em doubly degenerate}. In other words, we observe in this example (see Appendix on page \pageref{app} and also \cite{ss}) the
power of the mixed anomalies -- the prediction of the projective action of the
symmetries and the ground state degeneracy.
This is a strong result at strong coupling (i.e. at $eL\gg 1$). Sorry for the pun... After \cite{K1,K2,K3} a large number of non-trivial applications has been worked out.
Many relevant references can be found in \cite{da1,hkt}.

Other important advances in QFT/SC are connected with the approach which I would call  \"Unsal's continuity \cite{mitun} . Ten years ago Mithat \"Unsal and collaborators suggested the following strategy. If you want to analyze a certain Yang-Mills theory with a certain fermion sector, consider first this theory on $R_3\times S_1$ instead of $R_4$. The compactified direction of circumference $L$ can be viewed as spatial. If for small $L$ you manage to find boundary conditions that would guarantee unbroken center symmetry then there is no phase transition on the way to large (or infinitely large) $L$. The journey from weak to strong coupling confining phase is smooth. Hence all regularities observed at small $L$ remain valid also at strong coupling. This strategy gave rise to intriguing results, see e.g. \cite{MU}. In particular, it was found that {\em  adjoint} QCD exhibits a unique property of unexpected cancellations at large $N$ --  not only the leading contribution $\sim N^4$ in the vacuum energy  disappears but two subleading terms cancel as well! For one adjoint Weyl fermion this theory is supersymmetric, and cancellations in the vacuum energy are not surprising, of course. However, say, for two or three adjoint Weyl fermions there is no exact  supersymmetry, and yet the cancellation persists (although it is not exact in this case). 

The island of heavy quark physics exists and prosper. Experiments at  BESIII (Beijing),   Belle2  (Japan), and  LHCb (CERN) are active and produce lavish fruits. For instance, recently the lifetimes
of $b$-containing baryons have been precisely remeasured by LHCb. Disagreements (sometimes drastic) with the QCD-based predictions made in the 1980s and 1990s \cite{SV} are gone in the LHCb data,
the level agreement with theory is quite remarkable. But who cares? Heavy quark containing {\em pentaquarks} were found, and so on. If the ``old" pentaquarks were buried $\sim 10$ years ago, the new ones are here to stay. A suspicious violation of $\mu$-$e$ universality in semileptonic $B$ decays was reported recently. I believe it will go away with more precise measurements.
 
\vspace{2mm}

\centerline{***}

In the past, experimental data provided guidance for HEP theorists, and I think they will continue to do so. Experiment  and observation play the role 
of Polaris for courageous travelers at high seas
far from the shores. In the good old days, theorizing was like sailing between nearby islands of
experimental evidence. 
Now in search of hints of nature the HEP theorists have to travel deep in the ocean. Therefore, the nature of data they fish out may change.

\begin{figure}[h]
\epsfxsize=5.5cm
\centerline{\epsfbox{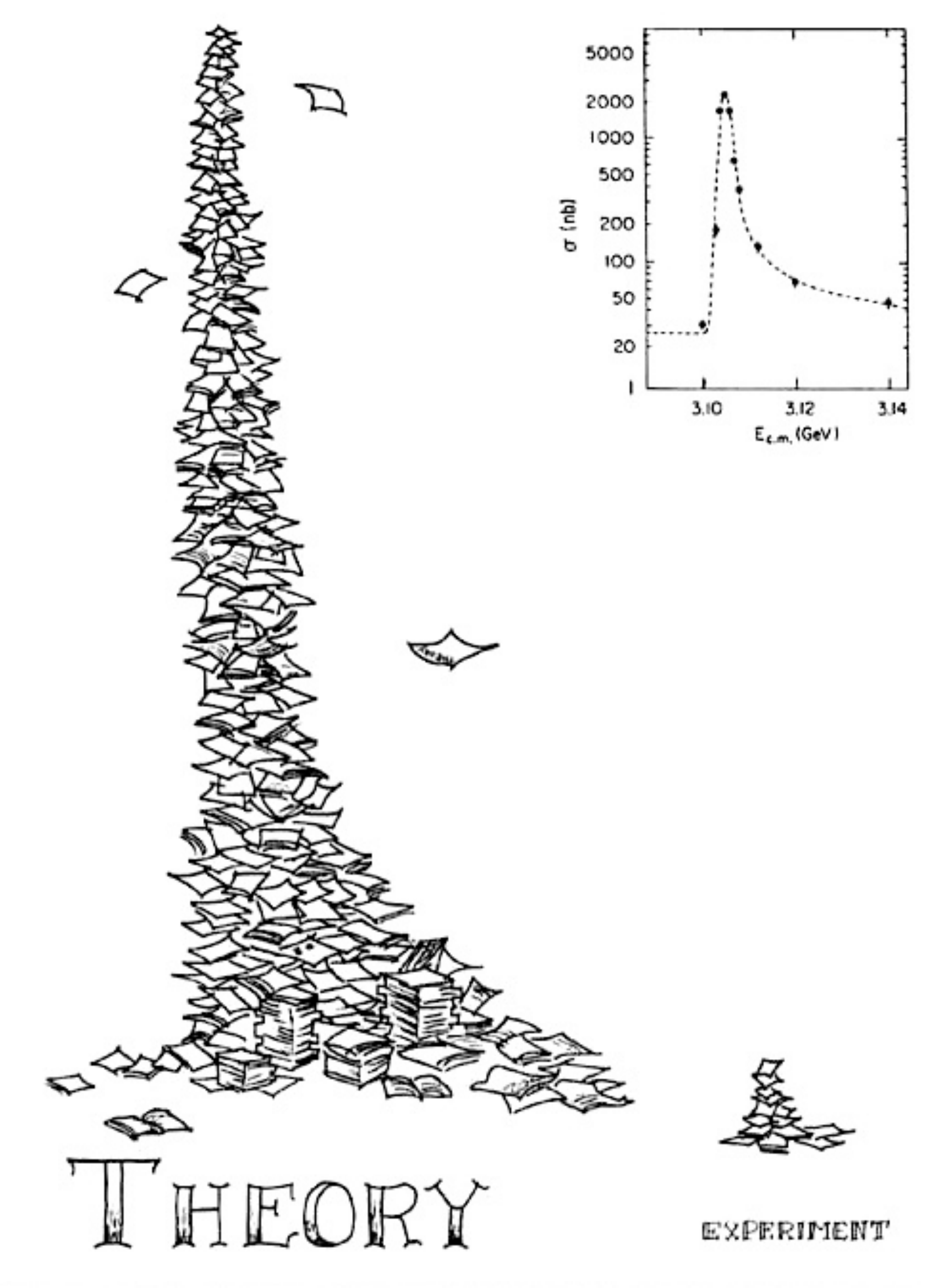}}
\caption{\small
Cartoon from  the cover of {\em CERN Courier,} April, 1975.}
\label{figu4ok}
\end{figure}

Experimental results were always less numerous (but way more precious) than theoretical production in the form of a stream of 
papers or conference talks. This was the case even in the glorious days following the November revolution in 1974 as is seen from the cartoon in {\em CERN Courier}, see Fig. \ref{figu4ok}.  

\vspace{2mm}

With increasing complexity of experiments and the need for more and more public funding it seems natural that the ratio exp/th would continue to fall in the near future.
The peak on the right may well  be shrinking for a while, while the peak on the left is growing unconstrained by rigors of nature.  This is a new scientific environment to which 
we, the physicists,  will have to adapt, as it usually happens in nature, through self-regulation. In the same way humankind adapts to new political and social conditions. In response to environmental changes populations grow or shrink. Theorists in their community are subject to the same social regularities. 

I hope that young people currently entering the area will focus more 
on the established mysteries of nature (e.g. dark matter) than in the 1990s' and early 2000s. I expect that on the way they will discover  new mysteries
(Fig. \ref{figu1ok}). Perhaps, they will 
come to a new scientific paradigm. Thinking boldly, why not imagine that quantum mechanics  gives place to something else at shorter distances? 
 Or a milder statement: why not replace the concept of naturalness by the following: As we move in the UV all interactions (including gravity) must remain weak? Perhaps, the information loss paradox in evaporating black holes might be solved, perhaps...
 
 I understand that uncovering  the fundamental laws of nature became harder due to scarcity of adequate  probes for experimentation.
 Does it mean that we have to give up right now?


\vspace{2mm}

\centerline{***}


\begin{flushright}
 \begin{minipage}[h]{20em}
{\em 
For never was a story of more woe \\
Than this of meta-induction show}. \cite{DS}
\end{minipage}
\end{flushright}

\vspace{1mm}

In 2013 Richard Dawid, a HEP theorist turned philosopher, published a book entitled ``String Theory and the Scientific Method" \cite{RD} which caused a significant resonance in the 
community.  This book was a response to the spread of the idea that from now on theories will not need empiric confirmation. It was based on the assumption that physicists' pursuit for quantum gravity
(through string theory) and early cosmology (through Multiverse) cannot be supported by data in principle, and instead, the emerging theory should be subject to the test of ``non-empirical confirmation."
According to Dawid, three principles of non-empirical confirmation are to replace experimental data/observations:

(i) The absence of alternatives in the community;

(ii) The degree to which a theory is connected to already confirmed theories (also referred to as meta-induction);

(iii) The amount of unexpected insights that the candidate ``non-empirically confirmed" theory gives rise to. 

I did not know what meta-induction was. Mathematical induction -- yes, but what's meta? I had to google this term and this is what I found in a philosophy dictionary and a number of articles (brief summary):
\begin{quote}
Epistemic optimism is a concept in which 
knowledge is perceived as the true representation of reality and science reveals what the world is. 
Meta-induction, (or, pessimistic induction) is an argument which seeks to rebut 
scientific realism, particularly the scientific realist's notion of epistemic optimism. Meta-inductive methods make predictions based on aggregating the predictions of different available prediction methods according to their success rates. The success rate of a method is defined according to some way of scoring success in making predictions, for instance, through the rate of approval in the community, especially its leading members.
\end{quote}

Shortly after, Dawid's book was criticized by George Ellis and Joe Silk in the article ``Scientific method: Defend the integrity of physics" \cite{ES} who defended the thesis 
``a theory must be falsifiable to be scientific." This triggered an ongoing debate in the community.

On December 7-9, 2015,  around 100 physicists and philosophers gathered in Munich at a conference provocatively entitled  ``Why trust a theory? Reconsidering Scientific Methodology in Light of Modern Physics." Among distinguished physicists one should note such esteemed theorists as Gia Dvali, David Gross, Dieter L\"ust, Slava Mukhanov, Joe Polchinsky and others.

The prevailing theme was that physics entered a new era of the so-called post-empirical science (PES), see a red murshroom at the right bottom in 
Fig.~\ref{figu3}. 
Starting from Galileo it was believed that the ultimate judge of any theory was observation and experiment. ``Not any longer," was the {\em leitmotiv} of many talks. 

With all due respect I strongly disagree with Richard Dawid and all supporting speakers at the conference and beyond. David Gross suggested a reconciling compromise. Here is a brief paraphrase of one of his statements:  ``It is only theories which need experimental confirmation, frameworks do not. The Standard Model is a theory, and it was 
triumphantly confirmed. But QM, QFT and ST are frameworks, not theories, they need not be confirmed in the usual way. With regards to frameworks, Dawid's criteria (i), (ii), and (iii) should be applied." 

David Gross is a great theoretical physicist, whose discovery of asymptotic freedom made him immortal, but I respectfully disagree with him. Framework or not\,\footnote{What's in a name? that which we call a rose
by any other name would smell as sweet \cite{DS}.} both QM and QFT have absolutely solid confirmations in all their aspects in thousands of experiments. I can agree to call them frameworks, alright, but I insist that  QM and QFT beyond any doubt describe our world at appropriate distances.  

I object against applying the term ``non-empirically confirmed" to science (the more so, the term ``postempiric science"). Of course, we live in liberal times and everybody is entitled to 
 study and discuss  whatever he or she wants. But the word science is already taken. Sorry, colleaugues.  For ``postempiric science," please, use another word, for instance,
iScience, xScience, or something else.

Even in such vague disciplines as, say, sociology
scholars search for empiric confirmation of their theories. The only exception is  mathematics to some extent, as I argued on page \pageref{p6}.

\begin{figure}[h]
\epsfxsize=11cm
\centerline{\epsfbox{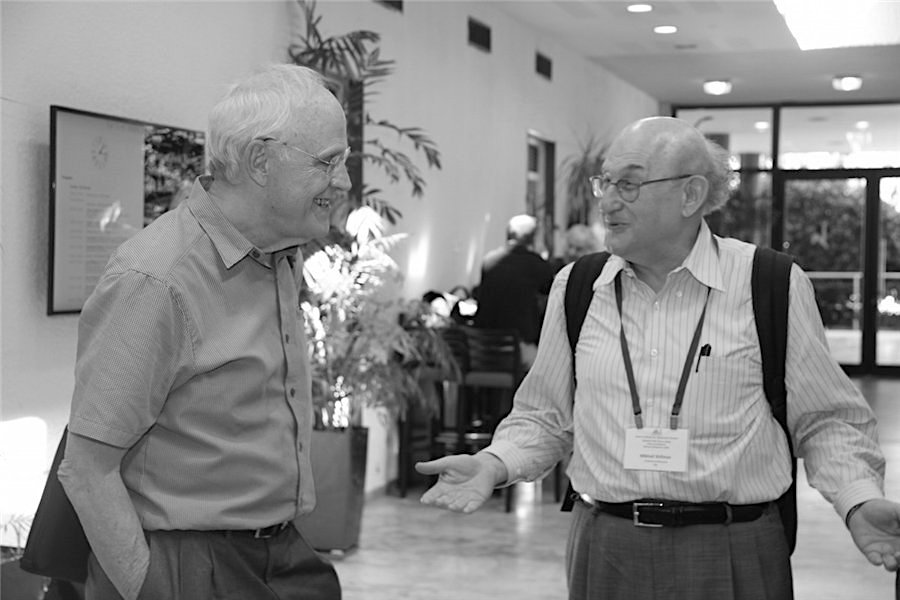}}
\caption{\small
A friendly discussion with David Gross.}
\label{figu4o}
\end{figure}

\label{p14}
Yes, I accept Multiverse and I even like this concept. After all, the Sun is not unique, our Galaxy is not unique, why our Universe should be unique? Multiverse  can be perceived as a poetic symbol or religion, which absolves us,\footnote{Through the Anthropic Principle.} Earth dwellers, from the need to explain hierarchies, in much the same way as observant Christians do not have to explain immaculate conception. But what is there for modern physicists? The word {\em physics} came from Ancient Greece.  In Greek $\tau \alpha \,\, \phi \upsilon \sigma \iota \kappa \alpha $ means the natural things. This was the title of Aristotle's treatise on nature.  In HEP this indeed used to be the case until recent years when  HEP's connection to our world started fading away with folding of large accelerator programs. However, neutrino physics is flourishing. Moreover, in physics at very large distances  observations are abundant, and in CM which deals with natural phenomena by definition the stream of data will not dry out in the foreseeable future. 
I believe that pause in accelerator programs we are witnessing now is not necessarily the same as the end of explorations at short distances. They will continue, perhaps in a new form, with novel devices,  and at a different pace.
Something will come up.
\vspace{2mm}

\centerline{***}


I am gtrateful to Alexey Cherman, Andrey Chubukov, Alexander Gorsky, Alexey Kamenev, Mikhail Katsnelson, Andrey Losev, Eric Poppitz, Mithat \"Unsal, and Zohar Komargodski for useful discussions. This work is supported in part by DOE grant DE-SC0011842.

\begin{flushright}
$\DAlambert$
\end{flushright}

\vspace{-7mm}

\begin{center}
\em \large
Appendix

\end{center}
\label{app}

In two-dimenional Schwinger model with one massless Dirac fermion of charge 2 and non-dynamical probe fermion of charge 1 with  compactified  spatial dimension (a circle of circumference $L$ with either periodic or antiperiodic boundary conditions on the fermion fields)  the Polyakov line along the compactified dimension can take two values
\beq
P = \left\langle   \exp\left( i \int_0^{L} dx A_1(x,t)\right)\right\rangle_{\rm ground\, st} = \pm 1\,,
\label{oneo}
\eeq
This corresponds to a $Z_2$ center symmetry. Note that the order parameter in $(\ref{oneo})$ is non-local, it is represented by a 1-form.

Another discrete symmetry in this example is the remnant of the U(1) chiral rotations. The U(1) axial symmetry in the Lagrangian is explicitly broken by the axial (diangle in the case at hand) anomaly,
but its discrete $Z_4$ subgroup survives \cite{shia}. This is due to the fact that the axial charge is conserved modulo 4, namely, $\Delta Q_5 = -4$. The latter circumstance implies in turn
that a four-fermion condensate $\langle\bar\psi \psi\rangle $ develops a nonvanishing expectation value. 
The chiral $Z_4$ symmetry is, in fact, $Z_2\times Z_2$, where the first factor is the fermion parity, i.e. $(-1)^F$, and the second factor is related to the sign ambiguity in the bifermion condensate 
$\langle\bar\psi \psi\rangle $ if it develops. A nonvanishing value of $\langle\bar\psi \psi\rangle $ would mean that the chiral $Z_4$ is broken down to $Z_2$.

A closer look shows that we deal with two global $Z_2$ symmetries whose generators do {\em not} commute with each other \cite{da1}. Thus, only one of these symmetries can be implemented, the other one must be spontaneously broken. In both cases the ground state is doubly degenerate.

\end{document}